\documentstyle[aps,prl,twocolumn,floats,psfig,epsfig]{revtex}

\begin{document}
\draft
\twocolumn[\hsize\textwidth\columnwidth\hsize\csname@twocolumnfalse\endcsname
\title{
The critical behavior of 2-d frustrated spin models with noncollinear order
}
\author{Pasquale Calabrese$\,^1$ and Pietro Parruccini$\,^2$}
\address{$^1$  Scuola Normale Superiore and INFN,
Piazza dei Cavalieri 7, I-56126 Pisa, Italy. 
}
\address{$^2$ 
Dipartimento di Fisica dell'Universit\`a di Pisa
and INFN, 
Via Buonarroti 2, I-56127 Pisa, Italy.
\\
{\bf e-mail: \rm
{\tt calabres@df.unipi.it},
{\tt parrucci@df.unipi.it}
}}

\date{\today}
\maketitle

\begin{abstract}
We study the critical behavior of frustrated spin models with noncollinear 
order in two dimensions, including antiferromagnets on a triangular lattice
and fully frustrated antiferromagnets. 
For this purpose we consider the corresponding $O(N) \times O(2)$ 
Landau-Ginzburg-Wilson~(LGW) Hamiltonian and compute the field-theoretic 
expansion to four loops and determine its large-order behavior. 
We show the existence of a stable fixed point for the physically relevant 
cases of two- and three-component spin models.
We also give a prediction for the critical exponent $\eta$ which is $\eta =0.24(6)$ and $\eta =0.29(5)$ for $N=3$ and $2$ respectively.
\end{abstract}


]

\section{INTRODUCTION.}
 
The critical behavior of frustrated spin systems with noncollinear or 
canted order
has been the object of intensive theoretical and experimental studies.
An important and debated 
issue is the nature of the universality of 
the phase transition (see, e.g., Refs. \cite{Kawamura-98,rev-01} 
for reviews on this issue in the three dimensional case).

In physical magnets, noncollinear order is due to frustration that may arise
either because of the special geometry 
of the lattice, or from the competition of different kinds of interactions.
Typical examples of systems of the first type are two-dimensional
triangular antiferromagnets (AFT), where magnetic ions are located at each 
site of a two-dimensional triangular lattice.
At the chiral transition, they can  be described by using short-ranged 
Hamiltonians for $N$-component spin variables 
defined on a triangular lattice as
\begin{equation}
{\cal H}_{\rm AFT} = 
     - J\,\sum_{\langle ij\rangle}  \vec{s}_i \cdot \vec{s}_j 
\label{latticeH}
\end{equation}
where $J<0$ and the sum is over nearest-neighbor pairs.

In these spin systems the Hamiltonian is minimized by 
noncollinear configurations, showing a 120$^o$ spin structure.
Frustration is partially released by mutual spin canting,
and the degeneracy of the ground-state is limited to global O($N$) 
spin rotations and reflections. 
As a consequence,  at criticality there is 
a breakdown of the symmetry from O($N$) in the high-temperature phase to 
O($N-2$) in the low-temperature phase, implying a matrix-like order parameter.
Frustration due to interactions may be realized in the
frustrated XY model on a square lattice, governed by the Hamiltonian:
\begin{equation}
{\cal H}_{\rm FXY} = 
     - J\,\sum_{\langle ij\rangle}  \cos (\theta_i- \theta_j-A_{ij}) 
\label{latticeH2}
\end{equation}
where $\theta_i$ is the angle of $ \vec{s}_i $ with a datum-direction,
$A_{ij}$ is the quenched vector potential and the sum is over nearest-neighbor
 pairs. 
The frustration is determined by the sum of $A_{ij}$ around a plaquette: 
in the fully frustrated case (FFXY) this sum is equal to $\pi$.
The large number of studies on this Hamiltonian is mainly motivated 
by the fact that it describes the statistical
properties of a superconducting Josephson junction in a transverse magnetic field.

Field theoretical (FT) studies of systems with noncollinear order are based on the
O($N$)$\times$O($M$) symmetric  Hamiltonian~\cite{Kawamura-98,Kawamura-88}
\begin{eqnarray}
{\cal H} = \int d^d x 
&& \left\{ {1\over2}
      \sum_{a} \left[ (\partial_\mu \phi_{a})^2 + r \phi_{a}^2 \right] 
+ {1\over 4!}u_0 \left( \sum_a \phi_a^2\right)^2 \right. \nonumber\\
&& \left. + {1\over 4!}  v_0 
\sum_{a,b} \left[ ( \phi_a \cdot \phi_b)^2 - \phi_a^2\phi_b^2\right]
             \right\},\label{LGWH}
\end{eqnarray}
where $\phi_a$ ($1\leq a\leq M$) are $M$ sets of $N$-component vectors.
We will consider the case $M=2$, that, for $v_0>0$,
describes frustrated systems with
noncollinear ordering such as AFT's.
Negative values of $v_0$ correspond to simple ferromagnetic or 
antiferromagnetic 
ordering, and to magnets with sinusoidal spin structures~\cite{Kawamura-88}.
The same LGW Hamiltonian has also been used in other problems
such as that of the phase transition of the dipole-locked A phase of helium 
three \cite{mermin-79}.

Even if the critical behavior of the non-collinear two dimensional XY 
frustrated systems has been 
the subject of many recent theoretical  and experimental \cite{bhj-xy} 
studies, there is still 
no definite conclusion about the nature of the phase transitions that occur 
in these systems. 
It can be shown that these models possess a twofold 
degenerate ground state according to the chiral degeneracy \cite{ms-85,Kawamura-98}; 
chirality is 
invariant under global spin rotation $SO(2)$, while it changes sign under 
global spin reflection $Z_2$. This situation leads to have an order parameter 
space (defined as a topological space isomorphic to the set of ordered states
\cite{mermin-79}) $V=Z_2\,\times \,S_1 \equiv Z_2\,\times \,SO(2)$ different 
from the one of the normal $XY$ model.

Monte Carlo simulations performed using the
AFT model \cite{ljnl-84,ms-85}, the FFXY model 
\cite{tj-83,lkg-91,rj-92,tk-90,gn-93,olsson-95,jpc-97,ll-98,bd-98,md-01} or 
other related models (some exotic realizations of the 
frustrated XY model are the Coulomb Gas 
with half integer charge \cite{tk-88,lee-93}, 
the coupled XY-Ising model \cite{gkln-91,lgk-91,ngk-95}, 
the 19-vertex version \cite{knkb-94,nijs-92}, the frustrated XY model 
with zig-zag coupling \cite{bg-97} and  the 
XXZ model \cite{cvct-98,ss-99,fckf-00,wh-00} 
for particular values of certain parameters)
have partially clarified the nature of the transition. 
A strongly debated question is whether there is  only one critical temperature 
$T_c$, in which both the $SO(2)$ and the $Z_2$ symmetries are 
simultaneously broken, or
there are two successive phase transitions  at critical temperatures
$T_{SO(2)}$ and $T_{Z_2}$. 
In the case of two transitions, even the order in which they occur and the 
numerical values of the critical exponents are very controversial~(in many 
papers \cite{olsson-95,ljnl-84,cvct-98} it is claimed that the higher 
temperature transition is a standard Ising one). 
The main problem is that, if two transitions exist,  they have 
very close critical temperatures and a precise detection of them 
by a Monte Carlo simulation is very difficult, in fact all the authors that
observe a single transition do not exclude the possibility of two very close
ones~(e.g. a recent upper limit for the difference of the 
critical temperatures
is $\Delta T \sim 5 \cdot 10^{-3}$~\cite{md-01}). We summarize all Monte Carlo
results in Table \ref{tabMC2} while in Table \ref{tabaltro} we report the 
results of other kinds of works.
At last we want to mention that in some studies \cite{foda-88} it has been argued 
that the
FFXY model is equivalent to a superconformal field theory with central 
charge $c=3/2$, but also this issue is not fully understood. 

\begin{table}[t]
\caption{Monte Carlo results for frustrated XY system. 
$n^o$ is the number of transitions observed in the simulation.
When two transitions are detected the subscript 1 and 2 are related to the 
first and to the second one. Where different exponents are reported we use 
scaling laws to obtain $\nu$ and $\eta$.}
\begin{tabular}{l|c|l}
Model[Ref.] & $n^o$ & Exponents\\
\tableline \hline
FFXY \cite{tj-83}  &              & $\alpha=0$\\ 
AFT \cite{ljnl-84} & 1 & $\beta =0.123(3),  \, \gamma =1.73(5)$ \\ 
AFT \cite{ms-85}   & 2 & $\alpha_1=0 ,\, \eta_2=1/4$ \\
FFXY \cite{rj-92}  & 2 & $\nu_1=0.898(3), \, \eta_1=0.216,$ \\
                   &   & $\sigma_2=0.3069, \, \eta_2=0.1915$ \\
FFXY \cite{olsson-95}&2& $\nu_1=1$ \\
CG \cite{tk-88}    & 1 &          \\
CG \cite{lee-93}   & 2 & $\nu_1 = 0.84(5), \, \eta_1 =0.26(4)$  \\
XY-Ising \cite{gkln-91}&1& $\nu = 0.82 , \, \eta=0.35$ \\
XY-Ising \cite{lgk-91}&1& $\nu = 0.84 , \, \eta=0.30$ \\
FFXY \cite{gn-93}  & 1 & $\nu=0.80(4),\,  \eta=0.38(3)$ \\
FFXY \cite{lkg-91} & 1 & $\nu=0.85(3), \, \eta =0.31(3)$ \\
FFXY tr \cite{lkg-91} & 1 & $\nu=0.83(4), \, \eta =0.28(4)$ \\
FFXY \cite{bd-98}  & 1 & $\nu=0.852(2), \, \eta=0.20(1)$ \\
FFXY tr \cite{ll-98}  & 2 & $\nu_1 = 0.83(1), \, \eta_1=0.25(2)$ \\
XXZ \cite{cvct-98} & 2 & consistent with Ising \\
FFXY \cite{md-01}  & 1 &  \\
FFXY \cite{tk-90}  & 1 & c=1.66(4), $\nu \sim 1$, $\eta=0.40(4)$\\
FXY zz \cite{bg-97}($\rho =1.5$)& 1 & $\nu=0.80(1), \, \eta  =0.29(2)$\\
FXY zz \cite{bg-97}($\rho =0.7$)& 1 & $\nu=0.78(2), \, \eta  =0.32(4)$\\
\end{tabular}
\label{tabMC2}
\end{table}

\begin{table}[b]
\caption{Other results for frustrated XY systems.} 
\begin{tabular}{l|l|l|l}
Model[Ref.] & Method &$n^o$ & Exponents\\
\tableline \hline
FFXY \cite{jpc-97} & pos. space RG & 2 & first non consistent with Ising\\
                   &               &   & second consistent with KT \\
19v \cite{knkb-94} & transfer matrix & 1 & $\eta=0.28(2), \, \nu=0.81(3)$  \\ 
                   &                 &   & $ c=1.55(3)$ \\
19v \cite{nijs-92} & transfer matrix & 1 & $\eta=0.26(1)$, \\ 
                   &                 &   & $\nu=1.0(1), \, c=1.50(5)$ \\
\end{tabular}
\label{tabaltro}
\end{table}

The critical behavior of the Heisenberg frustrated antiferromagnet
is clearer; many experiments 
\cite{bhj-h} agree with theoretical predictions
based on topological considerations \cite{km-84,kk-93,wea-94}, Monte Carlo 
simulations \cite{km-84,kk-93,sy-93,sx-95,wea-95,cadm-00,ss-99}, the
application of appropriate $NL\sigma$ models 
\cite{dr-89,adj-90,kawamura-91,awe-92,addj-93} and by the use of 
ERG \cite{tdm-00}. 
The order parameter space
of this model is the group of 3D rotations, in fact  
$O(3)/Z_2 \equiv SO(3)\equiv P_3$ \cite{km-84}.
Homotopy analysis shows that the system bears a topological stable 
point defect ($\pi_1(SO(3))=Z_2$, see for an exhaustive discussion 
\cite{km-84,Kawamura-98})
characterized by a two valued topological quantum number and exhibits a phase
transition driven by the dissociation of the vortices.
This analysis suggests the occurrence of a KT phase transition mediated by 
$Z_2$ vortices (while the standard KT transition for XY unfrustrated 
ferromagnet bears $Z$ vortices\cite{kt-73,mermin-79}): 
this observation strongly suggests that the
frustrated non-collinear magnets might exhibit a novel phase transition, 
possibly belonging to a new universality class \cite{km-84}.
A strong difference between the standard KT transition and the frustrated
Heisenberg model is that the last one presents a mass gap also in the low
temperature phase \cite{km-84,kk-93}. 
Monte Carlo simulations seem to confirm this scenario with the 
occurrence of a novel phase transition \cite{km-84,cadm-00,ss-99}, 
although some 
authors \cite{wea-94,wea-95} think that it is a standard KT transition.

The $NL\sigma$ model is not able to describe a defect mediated
transition \cite{zumbach-95}, 
in fact the perturbative $\beta$ function that we obtain from a  
$NL\sigma$ model for  a system which in the high temperature phase has a 
symmetry described by the group $G$ and in the low temperature phase has the
symmetry of $H$, subgroup of $G$, depends only 
on the local structure of the coset $G/H$. 
For the AFT model the order parameter space ($SO(3)$) is locally isomorphic to
the order parameter space of the $O(4)$ vector model (the four dimensional 
sphere $S^3$), but these two spaces have different first homotopy groups:
$$ \pi_1(SO(3))=Z_2, \qquad \pi_1(S^3)=0.$$ 
For this reason 
the AFT model and the $O(4)$ vector model have the same $\beta$
function \cite{addj-93}, even if topological excitations take to a 
different energy spectra.
When these excitations are activated, the $NL\sigma$ model approach 
leads to wrong predictions.
Instead the perturbative $\beta$ function of the LGW model 
depends only on the representation of $G$ and could be 
sensitive to the topological degrees of freedom. 
Let us emphasize that
there is no general consensus on the effectiveness of the perturbative treatment of 
the LGW model with these topological degrees of freedom. We can only 
mention the fact that the estimate for $g^*$ (the critical four point 
renormalized coupling) obtained in Ref. \cite{os-00} for the XY 
unfrustrated model from the five-loop perturbative expansion directly in two
dimensions, and the resummation of the same observable in the framework of 
the $\epsilon$ expansion at the order $\epsilon^3$ \cite{pv-98,pv-00}
are in perfect agreement with all the other results of non-perturbative methods~\cite{np}.

The LGW hamiltonian (\ref{LGWH}) has been extensively studied in the 
framework of $\epsilon=4-d$ expansion \cite{Kawamura-98,Kawamura-88,prv-01}, 
in the
$1/N$ expansion \cite{Kawamura-98,Kawamura-88,prv-01} and at fixed dimension  $d=3$ 
\cite{as-94,prv-00}. The existence and the stability properties of the fixed 
points depend on the number $N$ and on the spatial dimensionality (see for an
exhaustive discussion \cite{Kawamura-98,rev-01}). In the $\epsilon$ expansion, for 
sufficiently large values of $N$, there are four fixed points: the Gaussian one, the
$O(2N)$ ($v=0$) one and two new fixed points situated in the region $u,v>0$,
called chiral and antichiral. 
In fixed dimension $d=3$, for the physically relevant cases of $N=2,3$ \cite{prv-00},
the six-loop perturbative analysis shows the existence of the four fixed points
cited above, of which the chiral is the stable one.

In this paper we present a field-theoretic study based 
on an expansion performed directly in 
two dimensions, as proposed for the $O(N)$ models by Parisi\cite{Parisi-80}. 

The paper is organized as follows. In sec. II we derive the perturbative 
series for the renormalization-group functions at four loops and discuss the 
singularities of the Borel transform. The results of the analysis are 
presented in sec. III. In sec. IV we draw our conclusions and we discuss some
 future and unsolved issues.

\section{THE FIXED DIMENSION PERTURBATIVE EXPANSION IN TWO DIMENSIONS.}

\subsection{Renormalization of the theory.}
The fixed-dimension field-theoretical approach~\cite{Parisi-80} represents an 
effective procedure in the study of the critical properties in the symmetric 
phase of systems belonging to the $O(N)$ universality class (see e.g. 
Ref.~\cite{ZJ-book}). So the idea is to extend this procedure to frustrated models
where there are two $\phi^4$ couplings with different symmetry 
\cite{as-94,prv-00}. One 
performs an expansion in powers of appropriately defined zero-momentum quartic
couplings and renormalizes the theory by a set of zero-momentum conditions
for the (one-particle irreducible) two-point and four-point correlation 
functions:

\begin{equation}
\Gamma^{(2)}_{ai,bj}(p) = \delta_{ai,bj} Z_\phi^{-1} \left[ m^2+p^2+O(p^4)\right],
\label{ren1}  
\end{equation}
\begin{equation}
\Gamma^{(4)}_{ai,bj,ck,dl}(0) = 
Z_\phi^{-2} m \left[  
{u\over 3}\,S_{ai,bj,ck,dl}+ {v\over 6} \,C_{ai,bj,ck,dl}\right]
\label{ren2} 
\end{equation}
where $\delta_{ai,bj}= \delta_{ab} \delta_{ij}$ and
\begin{eqnarray}
S_{ai,bj,ck,dl}&=&\delta_{ai,bj}\delta_{ck,dl} + \delta_{ai,ck}\delta_{bj,dl} +\delta_{ai,dl}\delta_{bj,ck}\, , \nonumber \\ 
C_{ai,bj,ck,dl}&=&\delta_{ab}\delta_{cd}(\delta_{ik}\delta_{jl}+\delta_{il}\delta_{jk})+\delta_{ac}\delta_{bd}(\delta_{ij}\delta_{kl}+\delta_{il}\delta_{jk})\nonumber \\ 
&&+\delta_{ad}\delta_{bc}(\delta_{ij}\delta_{kl}+\delta_{ik}\delta_{jl})-2\, S_{ai,bj,ck,dl}.
\end{eqnarray}

The above relations  allow  to relate the second-moment mass $m$, and the 
zero-momentum quartic couplings $u$ and $v$ to the corresponding bare parameters $r$, $u_0$ and $v_0$ of the Hamiltonian (\ref{LGWH}).

In addition one introduces a renormalization condition for 
$\Gamma^{(1,2)}$, the one-particle irreducible
two-point function with an insertion of the operator $\case{1}{2}\phi^2$:
\begin{equation}
\Gamma^{(1,2)}_{ai,bj}(0) = \delta_{ai,bj} Z_t^{-1}.
\label{ren3} 
\end{equation}

From the perturbative knowledge of the functions $\Gamma^{(2)}$, $\Gamma^{(4)}$ 
and $\Gamma^{(1,2)}$ one can determine the RG functions 
as series of the two couplings $u$ and $v$.

The fixed points of the model are defined by the common zeroes of the $\beta$ 
functions, 
\begin{eqnarray}
\beta_u(u,v) = m \left. {\partial u\over \partial m}\right|_{u_0,v_0} ,\quad \beta_v(u,v) = m \left. {\partial v\over \partial m}\right|_{u_0,v_0}\label{bet} 
\end{eqnarray}
while the stability properties of these points are determined by the 
eigenvalues $\omega_i$ of the matrix $\Omega$:
\begin{equation} 
\label{omega}
\Omega = \left(\matrix{\displaystyle \frac{\partial \beta_u(u,v)}{\partial u}
 &\displaystyle \frac{\partial \beta_u(u,v)}{\partial v}
\cr & 
 \cr \displaystyle \frac{\partial \beta_v(u,v)}{\partial u}
& \displaystyle  \frac{\partial \beta_v(u,v)}{\partial v}}\right)\; .
\end{equation}
A fixed point with two positive eigenvalues is stable (totally attractive), 
and it determines the behavior of the system at criticality. The eigenvalues 
$\omega_i$ are connected to the leading scaling corrections, which go as 
$\xi^{- \omega_i}\sim |t|^{\Delta_i}$ where $\Delta_i = \nu \omega_i$, with 
few exceptions as the two dimensional Ising model~(see for a discussion Ref. 
\cite{cccpv-00}).

The RG-functions $\eta_{\phi}$ and $\eta_t$ are defined in the usual way:
\begin{eqnarray}
\eta_\phi(u,v) &=& \left. {\partial \ln Z_\phi \over \partial \ln m}
         \right|_{u_0,v_0}
= \beta_u {\partial \ln Z_\phi \over \partial u} +
\beta_v {\partial \ln Z_\phi \over \partial v} ,\label{eta1}
\end{eqnarray}
\begin{eqnarray}
\eta_t(u,v) &=& \left. {\partial \ln Z_t \over \partial \ln m}
         \right|_{u_0,v_0}
= \beta_u {\partial \ln Z_t \over \partial u} +
\beta_v {\partial \ln Z_t \over \partial v}.\label{eta2}
\end{eqnarray}

These functions are related to the critical exponents that can be obtained using the scaling laws in the following way:
\begin{eqnarray}
\eta = \eta_\phi(u^*,v^*),
\label{eta_fromtheseries} 
\end{eqnarray}
\begin{eqnarray}  
\nu = \left[ 2 - \eta_\phi(u^*,v^*) + \eta_t(u^*,v^*)\right] ^{-1},
\label{nu_fromtheseries}
\end{eqnarray}  
\begin{eqnarray}
\gamma = \nu (2 - \eta),
\label{gamma_fromtheseries} 
\end{eqnarray}  
where $(u^*,v^*)$ is the position of the stable fixed point.

\subsection{The four loop series.}
 In this section we report our results on the perturbative expansion of the 
RG functions (\ref{bet}), (\ref{eta1}) and (\ref{eta2}) up to four 
loops. 
The results are reported in terms of the rescaled couplings
\begin{equation}
u \equiv  {8 \pi\over 3} \; R_{2N} \; \bar{u},\qquad\qquad
v \equiv   {8 \pi\over 3} \;R_{2N} \; \bar{v} ,
\label{resc}
\end{equation}
where $R_N = 9/(8+N)$. We use these rescaled couplings, different from the 
usual ones \cite{prv-00}, because they have finite fixed point values in the 
limit $N\rightarrow \infty$.

The series are:
\begin{eqnarray}
\bar{\beta}_{\bar{u}} =&& -\bar{u} + \bar{u}^2 + {1-N\over (4+N)} \bar{u} \bar{v} - {1-N\over (8+2N)} \bar{v}^2 \nonumber \\
&&+  \sum_{i+j\geq 3} b^{(u)}_{ij} \bar{u}^i \bar{v}^j
\label{bu} \\
\bar{ \beta}_{\bar{v}} =&& -\bar{v} + {-6+N\over (4+N)}\bar{v}^2 + {6\over (4+N)} \bar{u} \bar{v} + \sum_{i+j\geq 3} b^{(v)}_{ij} \bar{u}^i \bar{v}^j
\label{bv}  \\
\eta_{\phi} =&& {1.83417\,(1+N)\over (8+2N)^2} \bar{u}^2 + {1.83417\,(1-N)\over (8+2N)^2} \bar{u}\bar{v} \nonumber \\ &&- {1.37563\,(1-N)\over (8+2N)^2} \bar{v}^2+ \sum_{i+j\geq 3} e^{(\phi)}_{ij} \bar{u}^i \bar{v}^j.
\label{etaphi}  \\
\eta_{t} =&& -{2\,(1+N)\over (4+N)} \bar{u}  -{(1-N)\over (4+N)}\bar{v} \nonumber + {13.5025\,(1+N)\over (8+2N)^2} \bar{u}^2\\&&+ {13.5025\,(1-N)\over (8+2N)^2} \bar{u}\bar{v}-{10.1269\,(1-N)\over (8+2N)}\bar{v}^2 \nonumber \\ &&+\sum_{i+j\geq 3} e^{(t)}_{ij} \bar{u}^i \bar{v}^j
\label{etat}
\end{eqnarray}
where \begin{equation}
\bar{\beta}_{\bar{u}}= {3\over 16\pi} \, R_{2N}^{-1}\beta_{u},\qquad\qquad
 \bar{\beta}_{\bar{v}}= {3\over 16\pi} \, R_{2N}^{-1}\beta_{v}.
\label{resc2}
\end{equation}
The coefficients  $b^{(u)}_{ij}$,  $b^{(v)}_{ij}$, $e^{(\phi)}_{ij}$ and 
$e^{(t)}_{ij}$  are reported in the Tables \ref{t1}, \ref{t2} and \ref{t3}.
Note that due to the  rescaling (\ref{resc2}), the matrix element of  $\Omega$ 
 are two times the derivative of $\bar{\beta}$ with respect to $\bar{u}$ 
and $\bar{v}$.

We have verified the exactness of our series by the following relations:
\begin{itemize}
\item
For $N=1$ the functions $\beta_{\bar{u}}$, $\eta_\phi$ and $\eta_t$ reproduce 
the corresponding functions of the standard $O(2)$ model \cite{os-00}.
\item
For $\bar{v}=0$, the functions $\beta_{\bar{u}}(\bar{u},0)$, 
$\eta_\phi(\bar{u},0)$ and $\eta_t(\bar{u},0)$ reproduce
the corresponding functions of the O($2N$)-symmetric model \cite{os-00}.

\item 
For $N=M=2$ one can transform the Hamiltonian (\ref{LGWH}) into two independent
XY-models via the transformation~\cite{Kawamura-88}
\begin{eqnarray}
\phi_{11}={{\phi}'_{11}-{\phi}'_{22} \over \sqrt{2}}, \qquad \phi_{12}={{\phi}'_{12}+{\phi}'_{21} \over \sqrt{2}}, \nonumber \\
\phi_{21}={{\phi}'_{21}-{\phi}'_{12} \over \sqrt{2}}, \qquad \phi_{22}={{\phi}'_{22}+{\phi}'_{11} \over \sqrt{2}},
\end{eqnarray}
and with the condition $v_0=-2u_0$. One can easily verify that, calling $x_0=u_0-v_0/2$ the new bare coupling, the RG function $\bar{\beta_{\bar{x}}}$ reduce to the one of the XY-model:
\begin{eqnarray}
\bar{\beta}_{\bar{x}}=\frac{5}{6}\,\bar{\beta}_{\bar{u}}(\case{3}{5}\,\bar{x}, - \case{6}{5}\,\bar{x})-\frac{5}{12}\,\bar{\beta}_{\bar{v}}(\case{3}{5}\,\bar{x}, - \case{6}{5}\,\bar{x}).
\end{eqnarray}

\begin{figure}[b]
\centerline{\psfig{height=6truecm,width=8.6truecm,file=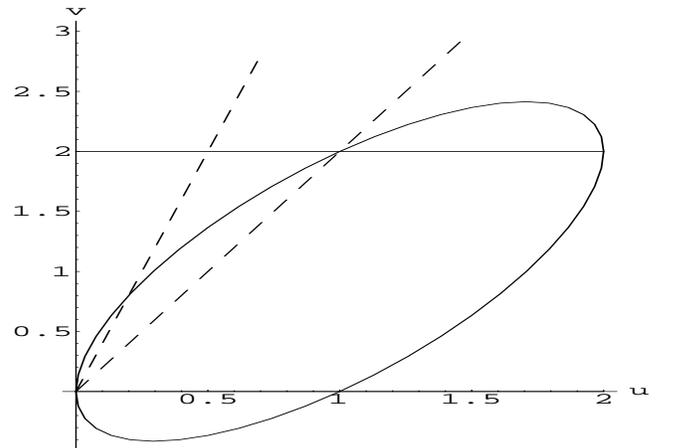}}
\caption{Zeroes of the $\bar{\beta}$-functions for $N=\infty$ in the 
$(\bar{u},\bar{v})$ plane. The two dashed lines represent the curves $z=2$ and $z=4$.}
\label{infinito}
\end{figure}

\item
In the limit $N \rightarrow \infty$ the RG functions reduce to:
\begin{eqnarray}
\bar{\beta}_{\bar{u}} =&& -\bar{u} + \bar{u}^2 -\bar{u}\bar{v}+\frac{1}{2}
\bar{v}^2\\
\bar{ \beta}_{\bar{v}} =&& -\bar{v}(1-\frac{\bar{v}}{2}) \\
\eta_{\phi} =&&0 \\
\eta_{t} =&& -2\bar{u}+ \bar{v}.
\end{eqnarray}
The zeroes of the two $\bar{\beta}$ functions are reported in Fig. \ref{infinito}.
There are four fixed points: the gaussian one (0,0) that is unstable with
$\omega_1=\omega_2=-2$, the $O(2N)$ symmetric one that is unstable in the
$v$ direction with $\omega_1=-\omega_2=2$, the chiral one that is the 
only stable with corrections to the scaling given by $\omega_1=\omega_2=2$ and
the antichiral fixed point that is unstable with the same eigenvalues of the 
$\Omega$ matrix of the $O(2N)$ one. 
We note that the region of attraction of the chiral fixed point is not all
the $(\bar{u},\bar{v})$ plane but only the region with $\bar{u}>0$ and $z=\bar{v}/\bar{u}<2$; if the system
is located out of this region it undergoes first order phase transition.
The critical indices of the gaussian and antichiral fixed points are  
mean-field~(i.e. $\eta=0$ and $\eta_t=0$), while the $O(2N)$ and the chiral
fixed point have non trivial exponents given by $\eta=0$ and $\eta_t=2$, as it is 
already known from large $N$ expansion~\cite{Kawamura-98}. 
This is another check of  exactness of the series.

\end{itemize}

\begin{table}[tbp]
\squeezetable
\caption{
Coefficients $b^{(u)}_{ij}$ 
of the four-loop expansion of 
$\bar{\beta}_{\bar{u}}$.
}
\renewcommand\arraystretch{0.3}
\begin{tabular}{c|l}
\multicolumn{1}{c}{$i,j$}&
\multicolumn{1}{c}{$R_{2N}^{-i-j} b^{(u)}_{ij}$}\\
\tableline \hline
3,0 &$ -0.52318301- 0.35762725\,N - 0.05670787\,{N^2} $ \\
2,1 &$ +0.27622342 + 0.20716756\,N + 0.06905585\,{N^2} $ \\
1,2 &$ +0.24421150 - 0.18315862\,N - 0.06105287\,{N^2} $ \\
0,3 &$ -0.06173989 + 0.04630492\,N + 0.01543497\,{N^2} $ \\ 
\hline
4,0 &$ +0.63938617 + 0.52357610\,N + 0.11532026\,{N^2} +0.00609697\,{N^3}$ \\
3,1 &$+0.52972245 - 0.34404956\,N - 0.17236232\,{N^2} -0.01331057\,{N^3}$ \\
2,2 &$-0.60971053+ 0.40445539\,N + 0.19204827\,{N^2} +0.01320688\,{N^3}$\\
1,3 &$+0.25988019- 0.18091027\,N - 0.075469945\,{N^2} -0.00349997\,{N^3}$\\
0,4 &$ -0.02523851 + 0.02158825\,N + 0.00431510\,{N^2}  -0.00066484\,{N^3}$\\
\hline
5,0 &$-1.02844874 - 0.96463436\,N - 0.27426277\,{N^2} 
 -0.02444185\,{N^3}$\\&$ - 0.00002407294\,{N^4}$\\
4,1 &$ -1.16733461 + 0.63874480\,N + 0.46960694\,{N^2} + 0.05903441\,{N^3} 
$\\&$ - 0.00005154248\,{N^4} $\\
3,2 &$+ 1.57188708 - 0.87883974\,N - 0.61913519\,{N^2} -0.07418884\,{N^3} 
$\\&$+ 0.0002766848\,{N^4} $\\
2,3 &$ -0.91454905 + 0.53282094\,N +0.34517685\,{N^2}+0.03698163\,{N^3}$\\
&$- 0.000430362\,{N^4} $\\
1,4 &$ +0.21807034- 0.13987590\,N - 0.07359806\,{N^2} - 0.00492708\,{N^3} $\\
&$+  0.0003307098\,{N^4} $\\
0,5 &$ -0.02243288 + 0.01659580\,N + 0.006205439\,{N^2}
-0.0002619636\,{N^3} $\\ &$ - 0.0001063929\,{N^4}$\\
\end{tabular}
\label{t1}
\end{table}

\begin{table}[tbp]
\squeezetable
\caption{
Coefficients $b^{(v)}_{ij}$ 
of the four-loop expansion of 
$\bar{\beta}_{\bar{v}}$.
}
\renewcommand\arraystretch{0.3}
\begin{tabular}{c|l}
\multicolumn{1}{c}{$i,j$}&
\multicolumn{1}{c}{$R_{2N}^{-i-j} b^{(v)}_{ij}$}\\
\tableline \hline
2,1 &$-1.01710219 - 0.38232322\,N- 
0.03201192\,{N^2}$ \\
1,2 &$ +0.76100686+ 0.17012363\,N -0.005032022\,{N^2}$ \\
0,3 &$ -0.17561980+ 0.00823505\,N +0.01303500\,{N^2}$ \\ 
\hline
3,1 &$+1.4980998 + 0.66351375\,N + 0.06331414\,{N^2} - 0.0022332645\,{N^3}$ \\
2,2 &$-1.49031358- 0.52140551\,N - 
0.012396398\,{N^2} + 0.006202595\,{N^3}$\\
1,3 &$+0.59692704+ 0.15592619\,N - 0.006944004\,{N^2} - 0.002154403\,{N^3}$\\
0,4 &$-0.07062074- 0.00996809\,N + 
    0.000144371\,{N^2} - 0.0004443507\,{N^3}$\\
\hline
4,1 &$-2.80757448 - 1.43132297\,N - 
0.19616848\,{N^2}- 0.004346583\,{N^3}$\\
&$ - 0.0002234496\,{N^4} $\\
3,2 &$+3.39501758+ 1.49420844\,N +0.13249155\,{N^2}$ \\
&$- 0.004523054\,{N^3} + 0.00067373429\,{N^4} $\\
2,3 &$-1.9481041- 0.78279460\,N- 0.05338075\,{N^2} + 0.002303844\,{N^3}$\\
&$-0.00070912608\,{N^4} $\\
1,4 &$+0.50147548 + 0.18762207\,N+ 0.01554819\,{N^2} +0.001191026\,{N^3} $\\&$ + 
    0.0002987006\,{N^4} $\\
0,5 &$ -0.04975526 - 0.01162418\,N - 0.000186500\,{N^2} - 0.000278729\,{N^3}$\\&$ - 0.000045297366\,{N^4}$\\
\end{tabular}
\label{t2}
\end{table}

\begin{table}[tbp]
\squeezetable
\caption{
Coefficients $e^{(\phi)}_{ij}$ and $e^{(t)}_{ij}$
of the four-loop expansion of 
$\eta_{\phi}$ and $\eta_t$.
}
\renewcommand\arraystretch{0.3}
\begin{tabular}{c|l}
\multicolumn{1}{c}{$i,j$}&
\multicolumn{1}{c}{$R_{2N}^{-i-j} e^{(\phi)}_{ij}$}\\
\tableline \hline
3,0 &$-0.00119855 - 0.00149819\,N -0.000299637\,{N^3} $ \\
2,1 &$-0.00179782 + 0.00134837\,N + 0.000449456\,{N^2} $ \\
1,2 &$+0.00179782 - 0.00157301\,N - 0.000224728\,{N^2} $ \\
0,3 &$-0.000599275 +0.000636729\,N - 
    0.0000374547\,{N^2} $ \\ 
\hline
4,0 &$ +0.00633062 + 0.00891668\,N + 0.00247304\,{N^2} - 
    0.000113013\,{N^3}$ \\
3,1 &$+0.01266124 - 0.00748913\,N - 0.00539814\,{N^2}+ 0.000226025\,{N^3} $ \\
2,2 &$-0.01549651 + 0.00967682\,N + 0.00598920\,{N^2} - 0.000169519\,{N^3} $\\
1,3 &$+0.00722529 - 0.00531519\,N - 0.00196661\,{N^2}+ 
    0.0000565063\,{N^3}$\\
0,4 &$-0.00105507 + 0.00116599\,N - 0.0000897290\,{N^2} - 0.0000211899\,{N^3}$\\
\tableline \hline 
\multicolumn{1}{c}{$i,j$}&
\multicolumn{1}{c}{$R_{2N}^{-i-j} e^{(t)}_{ij}$}\\
\tableline \hline 
3,0 &$-0.13266188 - 0.17878908\,N - 0.04612721\,{N^2} $ \\
2,1 &$-0.19899281 + 0.12980200\,N + 0.06919081\,{N^2}$ \\
1,2 &$+0.17738991 - 0.13415335\,N - 0.04323656\,{N^2} $ \\
0,3 &$-0.05120890 + 0.04833365\,N + 0.00287525\,{N^2} $ \\ 
\hline
4,0 &$+0.17299878 + 0.25743030\,N + 0.08514284\,{N^2} + 
0.000711326\,{N^3} $ \\
3,1 &$+0.34599757 - 0.17713453\,N - 0.16744038\,{N^2} - 
0.00142265\,{N^3} $ \\
2,2 &$-0.39410819 + 0.21062748\,N + 0.18241372\,{N^2} + 0.00106699\,{N^3}$\\
1,3 &$+0.16999405 - 0.10189909\,N - 0.06869231\,{N^2} + 
0.000597349\,{N^3}$\\
0,4 &$-0.02028992 + 0.01767450\,N + 0.00343506\,{N^2} - 0.000819639\,{N^3}$\\
\end{tabular}
\label{t3}
\end{table}

\subsection{Resummations of the series.}

The field theoretic perturbative 
expansion generates asymptotic series that must be resummed to extract the 
physical information about the critical behavior of the real systems.

Exploiting the property that these series are Borel summable for $\phi^4$ 
theories in two and three dimensions  \cite{bores}, one can resum these 
perturbative expressions using a Borel transformation combined with a 
method for the analytical
extension of the Borel transform. If the domain of analyticity of the Borel 
transform is
known, one can perform a mapping which maps the domain of analyticity (cut at
the instanton singularity) $g_b$ onto a circle, in order to have the  maximal 
analyticity.
In the case of the $O(N)$ symmetric model, let be $F(g)=\sum f_kg^k$ any Borel
summable function, whose expansion in series has to be resummed; exploiting 
the knowledge of the large order behavior of the coefficients $f_k$ \cite{ZJ-book}
\begin{equation}
f_k \sim k! \,(-a)^{k}\, k^b \,\left[ 1 + O(k^{-1})\right] \qquad
{\rm with}\qquad a = - 1/g_b,
\label{lobh}
\end{equation}
(a large order behavior related to the singularity $g_b$ of the Borel transform closest to the origin) one can perform the following mapping~\cite{L-Z-77}
\begin{equation}
y(g) = {\sqrt{1 - g/g_b} - 1\over \sqrt{1 - g/g_b} + 1 }
\end{equation}
to extend the Borel transform of $F(g)$ to all positive values of $g$ . 
The singularity $g_b$ depends only on the 
considered model and can be obtained from a steepest-descent calculation in which the relevant saddle point is a finite-energy solution (instanton) of the classical field equations with negative coupling~\cite{Lipatov-77,B-L-Z-77}. Instead the coefficient $b$ depends on which Green's function is considered.

Note that the function $F(g)$ can be Borel summable only if there are no
singularities of the Borel transform on the positive real axis.

This resummation procedure has worked successfully for the $O(N)$ symmetric 
theory, for which accurate estimates for the critical exponents and other 
physical quantities have been obtained. 

For this reason, the idea is to extend the resummation procedure cited above to
frustrated model, as it has been done for the three-dimensional case \cite{prv-00},
considering a double expansion in $\bar{u}$ and $\bar{v}$ at fixed $z=\bar{v}/\bar{u}$, and  studying the large order behavior (following the 
same procedure used in \cite{cpv-99,prv-00}) of the new expansion in powers of
$\bar{u}$ to 
calculate the singularity of the Borel transform closest to the origin 
$\bar{u}_b$.
The results are:
\begin{eqnarray}
{1\over \bar{u}_b} =& - a\, R_{2N}
\qquad & {\rm for} \qquad  0< z<4 ,
\label{bsing} \\
{1\over \bar{u}_b} =& - a\, R_{2N} \,\left( 1  - \displaystyle {1\over 2} z \right)
\qquad & {\rm for} \qquad   z<0 ,\quad z>4,
\nonumber
\end{eqnarray}
where $a = 0.238659217\dots$.

We  find that for $z>2$ there is a singularity on the real positive 
axis which  however is not the closest one to the origin for $z<4$. Thus, for
 $z>2$ the series are not Borel summable.

An important issue in the fixed dimension approach to critical phenomena 
concerns the analytic properties of the $\beta$ functions. 
As shown in Ref. \cite{cccpv-00} for the $O(N)$ model, the presence of 
confluent singularities in the zero of the perturbative $\beta$ function
causes a 
slow convergence of the resummation of the perturbative series to the correct
fixed point value. The $O(N)$ two-dimensional field-theory estimates of 
physical quantities 
\cite{L-Z-77,os-00} are less accurate than the ones of the three 
dimensional case, due to the stronger non-analyticities at the fixed point
\cite{cccpv-00,nickel-82,pv-98}.
In Ref. \cite{cccpv-00} it is shown that the non analytic terms cause
a large deviation in  the estimate of the right correction to the scaling,
i.e. the exponent $\omega$; instead the result for the fixed point value is a rather
good approximation of the correct one~(the systematic error is always less 
then 10\%). We think that this scenario holds also for frustrated models.

\section{FOUR-LOOP EXPANSION ANALYSIS.}
\label{3}
\subsection{The analysis method.}
The analysis of the four-loop series is performed following the 
procedure used in \cite{prv-00}: we exploit the knowledge of the value 
of the singularity of the Borel transform closest to the origin (a value given 
in the previous section), and we generate a set of approximants to our 
asymptotic series, varying the two parameters $\alpha$ and $b$.

If  
\begin{equation}
R(\bar{u},\bar{v}) =\, \sum_{k=0} \sum_{h=0} R_{hk} \overline{u}^h \overline{v}^k,
\end{equation}
is our asymptotic expression for one of the functions $\bar{\beta}$ or $\eta$, then our approximants can be written as:
\begin{eqnarray}
E({R})_p(\alpha,b;\bar{u},\bar{v})&=& \sum_{k=0}^p 
  B_k(\alpha,b;\bar{v}/ \bar{u}) \nonumber \\
 & \times &  \int_0^\infty dt\,t^b e^{-t} 
  {y(\bar{u} t;\bar{v}/\bar{u})^k\over [1 - y(\bar{u} t;\bar{v}/ \bar{u})]^\alpha}
\label{approx}
\end{eqnarray}
where
\begin{equation}
y(x;z) = {\sqrt{1 - x/\overline{u}_b(z)} - 1\over 
          \sqrt{1 - x/\overline{u}_b(z)} + 1}.
\end{equation}
The coefficient $B_k$ are determined by the condition that the expansion of 
$E({R})_p(\alpha,b;\bar{u},\bar{v})$ in powers of $\bar{u}$ and $\bar{v}$ gives $R(\bar{u},\bar{v})$ to order $p$.

In order to find the fixed points of the theory we  compute the series 
(\ref{approx}), with $R=\bar{\beta}_i$, for many values of $\alpha$ and $b$ (and for
several values of $N$), obtaining many different estimates of the $\bar{\beta}$ 
functions in all the plane $(\bar{u},\bar{v})$. We note that for $\alpha \geq 3$, 
the estimates strongly oscillates with varying $b$. For this reason 
we choose to keep $\alpha$ in the range $0 \leq \alpha \leq 2$. 
Then we look for the values of $b$, at fixed $\alpha$,  which make the 
series, at small values of $\bar{u}$ and $\bar{v}$ (i.e. $\bar{u}, \bar{v}\leq
 1$), more stable while varying the order $p$.
We realize that
all values of $b$ between 0 and 20 give reasonable estimates
of the $\bar{\beta}$ functions. For the estimates of the functions for larger
values of $\bar{u}$ and $\bar{v}$ we take the approximants in this range
of $\alpha$ and $b$.

At the end of this exploratory analysis, the strategy  
to obtain reasonable values and error bars for the fixed points, is to  
divide the domain $0\leq \bar{u}\leq 4$,
$0\leq \bar{v}\leq 6$ in $40^2$ rectangles, then to take 18 different approximants
with $b=5,7,9,11,13,15$ and $\alpha=0,1,2$ for the $\bar{\beta}$ functions, 
and to mark all the sites in which at least two approximants for 
$\bar{\beta}_{\bar{u}}$ and $\bar{\beta}_{\bar{v}}$ vanish. This procedure is
applied to the three-loop and four-loop series.

We adopt this method also for the values of $z$ for which the series are not
Borel summable. It should provide a reasonable estimate if $z<4$, because
we take into account the leading large order behavior.

\subsection{Instability of the symmetric fixed point.}

To begin with, we present the results related to
the instability of the $O(2N)$ fixed point. An accurate estimate of its 
location already exists from the Pad\'e-Borel analysis of the
five-loop series of Ref. \cite{os-00}. 
In table \ref{tab-inst} we report our estimates for 
$$\omega_v= 2 \left.{\partial \bar{\beta}_{\bar{v}}}\over{\partial \bar{v}} \right|_{\bar{u}=
\bar{u}_{O(2N)}^*,\bar{v}=0}$$
where $\bar{u}_{O(2N)}^*$ is the fixed point value obtained in \cite{os-00}.
The missing $\bar{u}_{O(2N)}^*$ values  are computed by the 
standard resummation procedure of \cite{L-Z-77}. 
These results clearly show that this fixed point is always unstable.

\begin{table}[t]
\caption{Crossover exponent $\omega_v$ at the $O(2N)$ fixed point.} 
\begin{tabular}{r|r|r}
N & $\bar{u}_{O(2N)}^*$ & $\omega_v$ \\
\tableline \hline
2 & 1.70(2)  &  -0.36(4)\\
3 & 1.62(1)  &  -0.7(2)\\
4 & 1.52(1)  &  -0.9(2)\\
6 & 1.407(3) &  -1.2(2)\\
8 & 1.313(3) &  -1.3(1)\\
16& 1.170(2) &  -1.70(4)\\
\end{tabular}
\label{tab-inst}
\end{table}

\subsection{Large ${\mathbf N}$ analysis (${\mathbf N\geq 4}$)}

\begin{figure}[b]
\centerline{\psfig{height=6truecm,width=8.6truecm,file=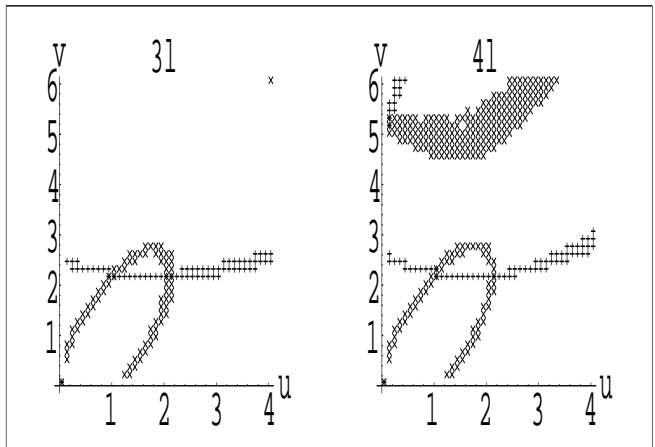}}
\caption{Zeroes of the $\bar{\beta}$-functions for $N=32$ in the $(\bar{u},\bar{v})$ plane.
Pluses ($+$) and crosses ($\times$) correspond to zeroes of $\bar{\beta}_{\bar{v}}(\bar{u},\bar{v})$ and $\bar{\beta}_{\bar{u}}(\bar{u},\bar{v})$ respectively.}
\label{n32}
\end{figure}

We begin our analysis of the four loop series from  large $N$ values, because
all these systems are free of topological defects and  they are well described 
by appropriate $NL\sigma$ models.
In fact, being the order parameter space 
$V\equiv O(N)/O(N-2)$, the first homotopy group is $\pi_1(V)=0$ for all 
$N\geq 4$~(this is a consequence of a simple theorem of differential
geometry, for which the $\pi_1(G/H)=0$ if $H$ is a continuous subgroup of a Lie Group
$G$).
We expect 
that, for these values of $N$, the chiral fixed point exists and it is stable with
critical indices and corrections to the scaling equal to the ones found for
$N=\infty$.

\begin{figure}[b]
\centerline{\psfig{height=6truecm,width=8.6truecm,file=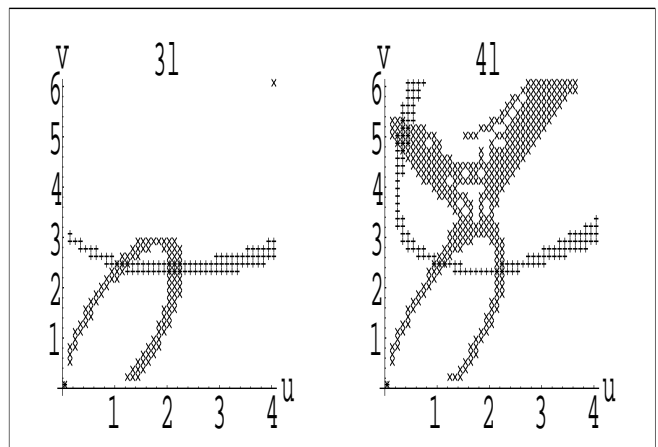}}

\caption{Zeroes of the $\bar{\beta}$-functions for $N=16$ in the $(\bar{u},\bar{v})$ plane.
Pluses ($+$) and crosses ($\times$) correspond to zeroes of $\bar{\beta}_{\bar{v}}(\bar{u},\bar{v})$ and $\bar{\beta}_{\bar{u}}(\bar{u},\bar{v})$ respectively.}
\label{n16}
\end{figure}

\begin{figure}[t]
\centerline{\psfig{height=6truecm,width=8.6truecm,file=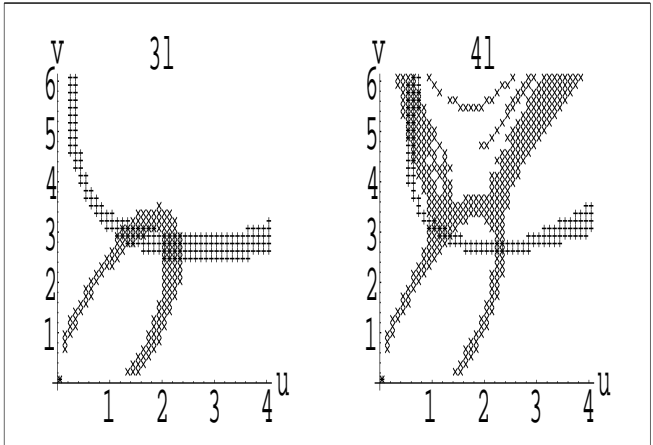}}
\caption{Zeroes of the $\bar{\beta}$-functions for $N=8$ in the $(\bar{u},\bar{v})$ plane.
Pluses ($+$) and crosses ($\times$) correspond to zeroes of $\bar{\beta}_{\bar{v}}(\bar{u},\bar{v})$ and $\bar{\beta}_{\bar{u}}(\bar{u},\bar{v})$ respectively.}
\label{n8}
\end{figure}

\begin{figure}[b]
\centerline{\psfig{height=6truecm,width=8.6truecm,file=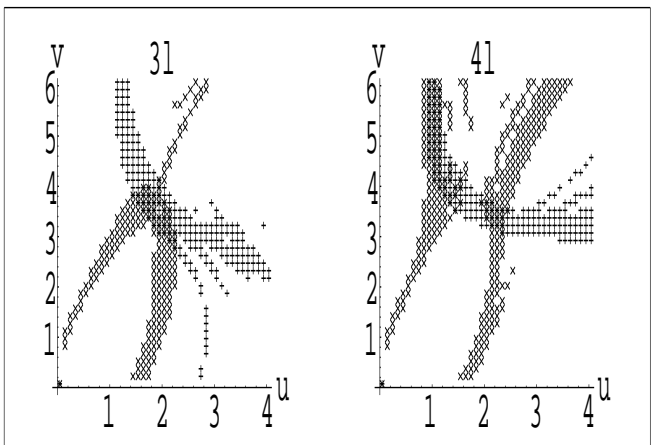}}

\caption{Zeroes of the $\bar{\beta}$-functions for $N=6$ in the $(\bar{u},\bar{v})$ plane.
Pluses ($+$) and crosses ($\times$) correspond to zeroes of $\bar{\beta}_{\bar{v}}(\bar{u},\bar{v})$ and $\bar{\beta}_{\bar{u}}(\bar{u},\bar{v})$ respectively.}
\label{n6}
\end{figure}

\begin{figure}[t]
\centerline{\psfig{height=6truecm,width=8.6truecm,file=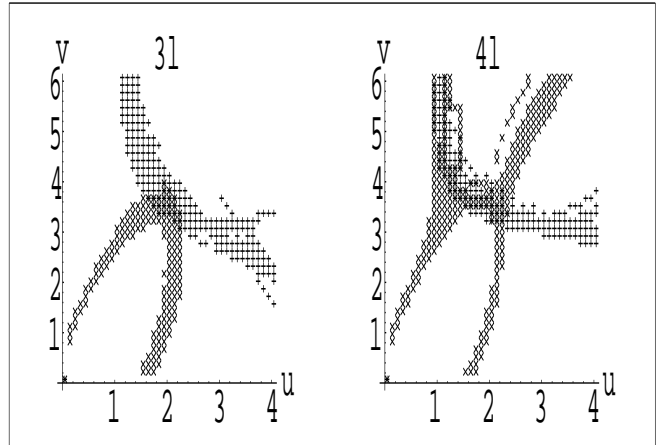}}

\caption{Zeroes of the $\bar{\beta}$-functions for $N=4$ in the $(\bar{u},\bar{v})$ plane.
Pluses ($+$) and crosses ($\times$) correspond to zeroes of $\bar{\beta}_{\bar{v}}(\bar{u},\bar{v})$ and $\bar{\beta}_{\bar{u}}(\bar{u},\bar{v})$ respectively.}
\label{n4}
\end{figure}

We apply the procedure illustrated above for $N$ equal to 32, 
16, 8, 6 and 4~(the choice of these particular values of $N$ will be clear in 
the follow). The results for the zeroes of the $\bar{\beta}$ functions, obtained from
the analysis of three and four-loop series are reported in Figs. 
\ref{n32}, \ref{n16}, \ref{n8}, \ref{n6}, \ref{n4}.

\begin{table}[t]
\caption{Fixed points  for $N>4$}
\begin{tabular}{r|c|c}
N & chiral fixed point~ &  antichiral fixed point\\
  &$(\bar{u}^*,\bar{v}^*)$& $(\bar{u}^*,\bar{v}^*)$\\
\tableline \hline
32 & (2.15(5),2.18(8))  &(1.05(5),2.25(15)) \\
16 & (2.2(1),2.40(15))  &(1.05(15),2.48(8)) \\
8  & (2.3(1),2.70(15))  & (1.15(25),3.08(22))\\
6  & (2.2(2),3.38(22))  & (1.2(3),$>$3.6) \\
\end{tabular}
\label{tabexp}
\end{table}

For $N=32$ the presence of the four fixed points predicted by the $1/N$ 
analysis is clear.
From the Fig. \ref{n32} emerges a new characteristic of
$\bar{\beta}_{\bar{u}}$: the appearance of a second branch of zeroes in the region of
upper values of $\bar{v}$.
This second branch exists only for the four-loop series, as in the three
dimensional case \cite{prv-00}, but it is present for almost all the
considered approximants. For $N=32$ this upper branch has no particular
importance, but it will become fundamental for lower values of $N$.

The evaluation of the $\Omega$ matrix shows that the chiral fixed point is 
stable with corrections to scaling given by $\omega_1=2.2(2)$ and $\omega_2=1.7(2)$.
The antichiral fixed point is unstable with crossover exponents $ \omega_1=1.7(2)$ and $\omega_2=-1.3(2)$. 
We use  a 
large set of approximants~(all those used for the $\bar{\beta}$s) 
to take into account the strong effect of non 
analyticities. In  this way we have a big error with respect to the 
one obtained by using only  stability criteria, but our estimates are very
close to the correct values $\omega_1=\omega_2=2$.

For $N=16$ the four fixed points of the $1/N$ expansion appear but the curve of zeroes
of $\bar{\beta}_{\bar{v}}$ cross the upper branch of the zeroes of $\bar{\beta}_{\bar{u}}$ in the
point $(0.4(2),5.0(3))$. This new fixed point has never been observed before. 
We call it antichiral II fixed point. We can not exclude the
possibility that this novel fixed point is a numerical artifact of the resummation 
procedure since it  belongs to the region with $z>4$, i.e. 
the region in which the singularity of the Borel transform closest to the 
origin is on the positive real axis. This problem is not so relevant 
because this fixed point is 
found to be unstable 
and it does not influence the critical behavior of the system.
Fig. \ref{n8} shows that the shape of the zeroes of the $\bar{\beta}$ 
functions for $N=8$ is similar to the one of $N=16$. Being the stability
properties the same for these two cases we conclude that
the critical behavior should be the same for all values of $N$ between 8 and 16.

For $N=6$ the two antichiral fixed points seem to coalesce, but 
probably this is a consequence of the large error bar of
 these zeroes. To solve this problem it is necessary
to know the $\bar{\beta}$ functions at higher order. However one eigenvalue of the
stability matrix, in all this wide zone, is  always negative, so this problem
does not influence our understanding of the critical properties of the model.
The chiral fixed point is always the only stable one, nevertheless the estimates of
the corrections to scaling are now quite different from the exact ones,
in fact we obtain $\omega_1=2.4(6)$ and $\omega_2=0.8(4)$ instead of 
$\omega_1=\omega_2=2$.
This is probably due to the non-analytic terms that, as shown in Ref. 
\cite{cccpv-00}, have larger effects when $N$ decreases.

In the case of  $N=4$~(see Fig. \ref{n4}), the exact number of the 
fixed points is not known and it is not clear if they are  located
on the upper or on the lower branch of zeroes of $\bar{\beta}_{\bar{u}}$.
From our homotopy analysis, we expect that
this model is connected with $N=\infty$ and as a consequence the stable zero 
should be
on the lower branch of $\bar{\beta}_{\bar{u}}$. 
Reasonable estimates of fixed points are not feasible from Fig. \ref{n4}.
Anyway, if we mark the sites in which at least 3, 4 etc. approximants for
the $\bar{\beta}$s vanish, we find two distinct regions located  
near $\bar{u}\sim 2$ and $\bar{u}\sim 1$. The existence of at least two fixed 
points is clear, but their estimate  is not possible.

Having established the existence of a stable fixed point, we compute the
critical exponents from the perturbative series reported in the previous 
section.
We consider a large number of approximants varying $\alpha$ and $b$ in the 
range $0 \leq \alpha \leq 5$ and $0 \leq b \leq 30$ and we choose for the 
final estimates the ones  more stable varying the number of loops. 
Varying $N$ we find that the exponent $\eta$ takes small values~(e.g. 
$\eta=0.040(2)$ for $N=32$ and $\eta=0.11(4)$ for $N=8$) which are close
to the ones found for the $O(N)$ models for $N \geq 3$, but different from the
exact one $\eta=0$, expected for asymptotic free theory. 
We think that these erroneous predictions are due to 
non-analytic terms of $\eta(u,v)$ at $(u^*,v^*)$, as for the $O(N)$ models.

\subsection{Frustrated Heisenberg model}
\begin{figure}[b]
\centerline{\psfig{height=6truecm,width=8.6truecm,file=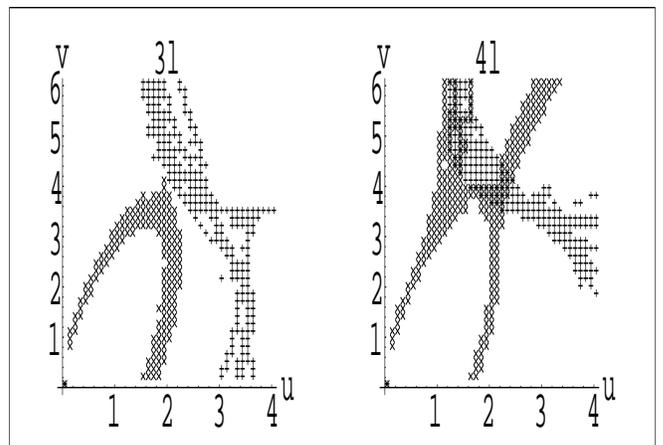}}

\caption{Zeroes of the $\bar{\beta}$-functions for $N=3$ in the $(\bar{u},\bar{v})$ plane.
Pluses ($+$) and crosses ($\times$) correspond to zeroes of $\beta_{\bar{v}}(\bar{u},\bar{v})$ and $\bar{\beta}_{\bar{u}}(\bar{u},\bar{v})$ respectively.}
\label{n3}
\end{figure}

The zeroes of the $\bar{\beta}$ functions for $N=3$ are reported in Fig. \ref{n3}.
It is apparent that the chiral fixed point is located on the upper branch of
zeroes of $\bar{\beta}_{\bar{u}}$, so this value of $N$ is not connected with 
$N=\infty$. For this reason the exponent $\eta$ could be different from zero.
We have already stressed the strong evidence for a topological 
phase transition for this frustrated system and so the physical reason for this
non trivial exponent is clear. 

The standard structure involving only four common zeroes of the $\beta$ functions 
is restored since we find only one antichiral fixed point. 
The chiral fixed point is located at 
\begin{eqnarray}
 \bar{u}^*=2.3(3) \qquad \qquad \bar{v}^*=3.9(5).
\end{eqnarray}
The eigenvalues $\omega_i$ vary  significantly with the two parameters 
$\alpha $ and $b$ and turn out to be complex in several cases. Reasonable 
estimates of $\omega_i$ are not feasible, nonetheless the sign of their
real part is positive for the majority of the approximants. 

The critical behavior of this topological phase transition is expected to be 
of the KT type, i.e. the correlation length and the susceptibility diverge,
for $t>0$ according to the asymptotic law:
\begin{equation}
 \xi \sim e^{b t^{-\sigma}}, \qquad \qquad \chi \sim \xi^{2-\eta}
\end{equation} 
where $\eta$ is the standard exponent. This means that we can not define the conventional exponents $\nu$ and $\gamma$ since $\xi$ diverges faster than any power of $t$.

For the $O(2)$ unfrustrated model Kosterlitz and Thouless \cite{kt-73} have 
shown that $\eta=1/4 $ and $\sigma =1/2$.
Exact values do not exist for the Heisenberg frustrated model.
The result of a very recent and accurate Monte Carlo simulation \cite{cadm-00} 
is $\sigma \sim 0.600$, even if  the standard KT form of 
thermodynamic quantities has been used in the past to fit Monte Carlo data 
finding a reasonable agreement \cite{kk-93,wea-95,sx-95}.

Our estimate for $\eta$ is:
 \begin{equation}
\eta=0.24(6).
\end{equation}
Note that this value is  consistent with the standard KT scenario.

This value of $\eta$, which is very different from zero,   seems to 
confirm that the LGW model is able to describe a topological phase 
transition \cite{nota}. 
We have already said in the introduction that there is no general consensus on
this issue. We think that this nontrivial result is a possible evidence 
in favor this thesis.

\subsection{Frustrated XY model}

As discussed in the introduction, the frustrated XY model breaks two
symmetries. If there are two transitions we are only able to 
describe the one that occurs at  higher temperature, since
we explore the critical properties of the model in the
high temperature phase.

\begin{figure}[t]
\centerline{\psfig{height=6truecm,width=8.6truecm,file=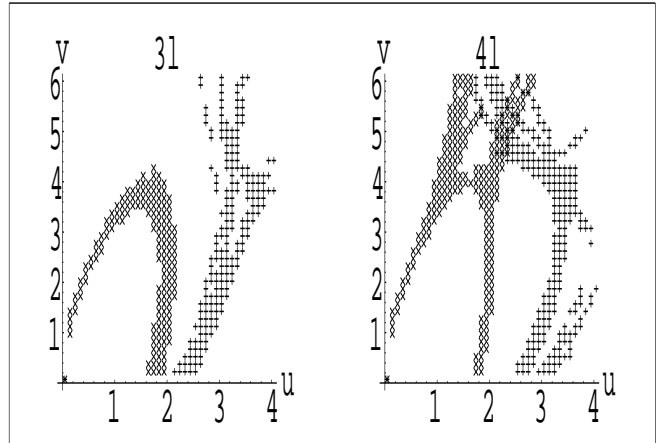}}
\caption{Zeroes of the $\bar{\beta}$-functions for $N=2$ in the $(\bar{u},\bar{v})$ plane.
Pluses ($+$) and crosses ($\times$) correspond to zeroes of $\bar{\beta}_{\bar{v}}(\bar{u},\bar{v})$ and $\bar{\beta}_{\bar{u}}(\bar{u},\bar{v})$ respectively.}
\label{n2}
\end{figure}

The Fig. \ref{n2} shows the zeroes of the $\bar{\beta}$ functions. 
The chiral fixed point is located at:
\begin{eqnarray}
\bar{u}^*=2.3(2),\qquad \bar{v}^*=5.0(5).
\end{eqnarray}
In the region that is not shown ($v>6)$ we find another fixed point that 
can be identified with  the antichiral.

To describe the critical behavior of these systems we have studied the 
stability of these two fixed points, evaluating the eigenvalues of the $\Omega$ 
matrix defined in eq.(\ref{omega}) and varying $\alpha$ and $b$ as explained
above. 
For the shortness of our perturbative series the results obtained are quite 
unstable to present a reasonable estimate,
 anyway we find that one of the two eigenvalues calculated at the antichiral 
fixed point is always negative, while for the chiral one the 
majority of the results are in agreement with its stability.

The evaluation of  the critical exponent $\eta$ at the chiral fixed point gives
\begin{equation}
\eta=0.29(5).
\end{equation}
This value is consitent with many Monte Carlo reported in Table \ref{tabMC2}.
We note that this estimate is substantially higher than $0.146$ obtained by the Pad\'e-Borel 
analysis of five-loop series for the
one-component $\phi^4$ theory \cite{os-00}.
We can not give a reasonable value for the exponent $\nu$ because of the big
oscillations observed when varying the two parameters $\alpha$ and $b$. We think that
longer series could solve this problem and partially clarify the nature of 
this phase transition.

\section{CONCLUSIONS}

In the present paper we have studied the critical behavior of
frustrated spin models with non-collinear order by applying for the first time 
the field-theoretic renormalization-group technique directly in two dimensions.

We have begun our analysis from large values of $N$, since these
systems are free of topological defects, and they are well described by
appropriate $NL\sigma$ models. We have considered $N= \infty$, 32, 16,
8, 6, 4, always finding the presence of the chiral and antichiral fixed
points. In order to verify the stability properties of these fixed points we
have evaluated the eigenvalues of the matrix $\Omega$, which demonstrate
the full stability of the chiral fixed point. We have also shown the
non-stability of the $O(2N)$ fixed point for $N \geq 2$, computing the 
crossover exponent $\omega_v$.

Then we have focused our attention on the frustrated $XY$ model ($N=2$)
and on the Heisenberg model ($N=3$), which are expected to undergo
continuous phase transitions that can not be described by the use of
$NL\sigma$ models.

For the XY model we have found the presence of a stable fixed point
located in $(2.3(2),5.0(5))$, in which the critical exponent $\eta$
takes the value 0.29(5). 

Even in the case of the frustrated Heisenberg model, we have found the existence of
a stable fixed point in   $(2.3(3),3.9(5))$. We have evaluated the
critical exponent $\eta =0.24(6)$.
This non trivial result confirms the validity of the LGW approach in 
describing defect mediated transitions.

The issue of the universality class of these systems may be clarified by 
the computation of the universal quantities $\bar{u}^*,\bar{v}^*$, 
related to 
the four point renormalized coupling constants, with a Monte Carlo simulation. 
In fact these quantities, which have been estimated in this work,
have finite values at the critical point and a precise determination of them may be more simple than the evaluation of  exponents.

\section*{ACKNOWLEDGMENTS.}
We would like to thank Ettore Vicari and Paolo Rossi for a critical reading 
of this manuscript and many useful discussions. 
We also thank M. Mintchev and A. Pelissetto for interesting discussions.



\begin{references}
\bibitem{nota}
We have resummed the four-loop series for the $O(3)$ model obtaining 
$\eta=0.11$ which is substantially lower than the value found in text for
the frustrated model.
\end{references}
\end{document}